\documentclass[a4paper,fleqn,usenatbib]{mnras}

\usepackage{ae,aecompl}

\usepackage{graphicx}	
\usepackage{amsmath}	
\usepackage{amssymb}

\title[Iron lines of inhomogeneous accretion flow]
{The Profiles of Fe K$\alpha$ Line From the Inhomogeneous Accretion Flow}

\author[X.-D. Yu et al.]{
Xiao-Di Yu,$^{1,3}$
Ren-Yi Ma$^{1,3,4}$\thanks{E-mail: ryma@xmu.edu.cn}
Ya-Ping Li,$^{2,3}$
Hui Zhang,$^{2,3}$
and Tao-Tao Fang$^{1,3}$
\\
$^{1}$Department of Astronomy and Institute of Theoretical Physics and Astrophysics, Xiamen University, Xiamen, Fujian 361005, China\\
$^{2}$Key Laboratory for Research in Galaxies and Cosmology, Shanghai Astronomical Observatory, \\ Chinese Academy of Sciences, 80 Nandan Road, Shanghai 200030, China\\
$^{3}$SHAO-XMU Joint Center for Astrophysics, Xiamen, Fujian 361005, China \\
$^{4}$Department of Astronomy, University of Massachusetts, Amherst, MA 01003, USA
}

\date{Accepted XXX. Received YYY; in original form ZZZ}

\pubyear{2015}

\begin{document}
\label{firstpage}
\pagerange{\pageref{firstpage}--\pageref{lastpage}}
\maketitle

\begin{abstract}
The clumpy disc, or inhomogeneous accretion flow, has been proposed to explain the properties of accreting black hole systems.
However, the observational evidences remain to be explored.
In this work, we calculate the profiles of Fe K$\alpha$ lines emitted from the inhomogeneous accretion flow through the ray-tracing technique,
in order to find possible observable signals of the clumps.
Compared with the skewed double-peaked profile of the continuous standard accretion disc,
the lines show a multi-peak structure when the emissivity index is not very steep.
The peaks and wings are affected by the position and size of the cold clumps.
When the clump is small and is located in the innermost region, due to the significant gravitational redshift, the blue wing can overlap with the red wing of the outer cold disc/clump, forming a fake peak or greatly enhancing the red peak.
Given high enough resolution, it is easier to constrain the clumps around the supermassive black holes than the clumps in stellar mass black holes due to the thermal Doppler effect.
\end{abstract}


\section{Introduction}\label{intro}
Based on the assumption of continuous fluid, different accretion modes have been established and are known as the standard accretion disc \citep[SAD;][]{SS73}, the advection-dominated accretion flow \citep[ADAF;][]{NY94,YN14}, and the slim accretion disc \citep[]{A88}. These models are widely accepted and are used to explain the properties of many kinds of sources in active galactic nuclei (AGN) and black hole X-ray binaries (BHXBs).
However, the real accretion flow may be more complicated,
especially when the accretion rate is close to the critical rate of mode transition.
One possibility is that cold clumps are formed in the hot gas, and the accretion flow becomes inhomogeneous.

Analytic works have shown that this inhomogeneous accretion flow can form for different reasons.
Considering the relative rate of evaporation and condensation, it is possible for the hot phase flow to condense in the inner region, and thus an interrupted cold disc forms \citep[]{ML07,L07,L11,QL12}.
Moreover, the hot phase flow could collapse onto the equatorial plane and form optically thick cold discs or clumps, when the accretion rate increases to a critical value, where the radiative cooling rate is stronger than the local heating rate caused by viscosity and compression work \citep[]{Y01,Y03,XY12}.
Additionally, instabilities such as the thermal \citep[]{K98}, magneto-rotational \citep[]{BS01,BS03}, or photon bubble \citep[]{G98,B02} instabilities can induce the clumps,
and the magnetic fields can play a role in maintaining the clumps \citep[]{C99}.
Recent numerical simulations have shown the cold clumps can survive in hot gases \citep[]{T02,T03,T04,W16,S17}.

Many works have shown that inhomogeneous accretion flow can explain the observations of BHXBs, AGNs and ultra-luminous X-ray sources (ULXs) more naturally.

In the hard state of some BHXBs, the relativistically broadened Fe K$\alpha$ lines and the component of thermal radiation have been observed \citep[]{M06a,M06b,R07,T08,R09,C10,R10,RN10,M15}.
These observations indicate that the cold disc exist in the innermost region,
but in the hard state, the inner accretion flow is usually thought to be the fully ionized hot gas.
A natural explanation is that cold clumps or an interrupted discs exist in the hot gas.

Although the bluer-when-brighter trend of AGNs could be explained by the changing of global accretion rate \citep[]{LC08,S11,GL13},
on the basis of the difference spectra of 604 variable quasars that have repeated observations in the Sloan Digital Sky Survey-I/II (SDSS),
\citet[]{R14} showed that the trend could not be reproduced by the global fluctuation of accretion rates alone.
They found the spectral variability could be reproduced by the inhomogeneous disc model with a large fluctuation in local temperature.
Moreover, if the characteristic timescales of fluctuations in the discs is radius-dependent \citep[]{DA11}, the timescale dependence of the colour variability in some quasars \citep[]{S14} could also be well reproduced \citep[]{C16}.

The soft X-ray spectra of some extreme ULXs show a thin disc like component and a roll-over in the 5-10keV band \citep[e.g.][]{M14}.
The origin of the roll-over is rather controversial.
Tf it is due to the Comptonization of low-temperature photons, an unreasonable corona is demanded,
with the temperature being $kT_e \simeq 2$~keV and optical depth being $\tau \simeq 10$ \citep[]{ST06,G09}.
In contrast, the patchy disc with a multiple temperature profile can explain the observed spectra without any unreasonable demanding \citep[]{M14}.

Some works have been devoted to exploring the theoretical properties of the inhomogeneous accretion flow.
The spectral and timing properties of the continuum have already been studied \citep[]{GR88,MC02,MM06}.
The physical properties of cold clumps under thermal and radiative equilibrium conditions were examined by \citet[]{K98}.
The dynamics of clumps embedded in ADAF was investigated by \citet[]{W12}.
Moreover, \citet[]{XY12} estimated the radiative efficiency of two-phase accretion flow and found that the radiative efficiency of the two-phase accretion flow could be as high as that of a standard disc.

In addition to the continuum, the Fe K$\alpha$ fluorescent line is another special probe for us to investigate the physical conditions around the black holes.
Although, as mentioned above, broad iron line in the hard state has indicated the existence of cold clumps in the inner region of the accretion flow,
further detailed investigations are still needed.
In this paper, based on a phenomenological model, we calculate the line profile and investigate the influences of the clump parameters. The model and method are introduced in Section \ref{mod}, the influences of the model parameters are investigated in Section \ref{res}, discussion and summary are given in Section \ref{dis} and \ref{con}, respectively.

\section{Toy Model}\label{mod}

\subsection{Geometry of the disc}\label{modd}
The inhomogeneous accretion flow discussed in this paper consists of three components: the hot gas, i.e., corona or ADAF, in the inner region, the outer SAD that is truncated by the hot gas at radius $r_{\rm t}$ and extends to the outer boundary of the accretion flow $r_{\rm out}$,
and the clumps of cold gas that are scattered within the hot gas.
The clumps lie in the range from $r_{\rm in}$ to $r_{\rm t}$, where $r_{\rm in}$ is the possible innermost radius that the clumps can exist.
As the SAD only extends inwards to innermost stable circular orbit (ISCO), we assume $r_{\rm in}$ to be the radius of ISCO.
The hard X-rays or ionization photons are produced in the hot gas,
and the fluorescent Fe K$\alpha$ line photons come from the cold gas, i.e., the clumps and the outer cold disc.

In this paper, we assume the clumps lie on the equatorial plane and rotate around the central black hole with Keplerian velocity like SAD.
Due to the tidal force and differential rotation,
clumps should be stretched and take the form of long arcs,
so we simplify them as a series of concentric rings being separated by the gaps of hot gas.
If we concentrate on the time-averaged line profile, this simplification could be relaxed as described in the section of discussion.
Each clump and its adjacent hot gas constituting a clump-gap unit.
We assume the units to be evenly distributed along the radius, i.e., all units are of the same width.
Although this assumption is simple, it is enough to predict the main features of the lines.

Without the knowledge of the number and size of clumps, we parameterize them as $n$ and $f$, respectively.
The parameter $f$ is the ratio of the width of the clump to that of the unit.
Since a clump occupies either the inner part or the outer part of the corresponding unit, the covering factor $f$ is set to be positive or negative, respectively.
When the number of clumps is small, the sign of $f$ leads to a great difference.
For example, when $n=1$, negative $f$ means the clump is connected with the outer SAD, and could be regarded as part of the outer SAD.
In this case, the increasing absolute value of $f$ corresponds to the process that $r_{\rm t}$ gradually recedes to ISCO.
While positive $f$ with increasing absolute value corresponds to the process that
an inner cold disc at ISCO gradually expands to $r_{\rm t}$.
For large $n$, the widths of the clump-gap units are small, and therefore the sign of $f$ does not matter much to the line profile.

Since the physical conditions of the source of hard X-rays are still uncertain,
the consequent emissivity of emission lines is assumed to be the power law of radius, i.e., $I(r) \sim r^{-p}$,
where the emissivity index $p$ could be a function of radius, depending on the geometry of hard X-ray sources \citep[]{WF12,D13}.
For simplicity we assume it to be constant in this paper.

\subsection{Ray-tracing technique}\label{modr}

The general relativistic effects on the emission lines can be calculated by following the trajectories of photons.
We can start either from the disc or from the image plane of observer and then find the other ends of the photons.
For given position of emission, the velocity, gravitational redshift and local emissivity of the iron line could be obtained.
And then we can calculate the observed frequency and intensity on the image plane.
The observed frequency could be described by the frequency shift factor $g\equiv \nu'/\nu_0$, where $\nu'$ and $\nu_0$ are the observed frequency and the frequency at the local rest frame, respectively.
After integrating over the disc or image plane, we can obtain the final profile of the relativistic line.
Many papers contributed to this method \citep[e.g.,][]{C75, L91,RB94, F97, C98, AK00, RN03, MC04, SB04, L05,YW13,Zhang15}.
In this paper we trace the photons on the image plane along their trajectories to the positions of emission on the equatorial plane \citep[]{RB94, F97, C98, MC04, SB04, L05}.
According to previous studies, both the black hole spin $a$ and the inclination angle $i$ of the line of sight to the axis of the disc are important to the line profile.

\begin{figure}
\begin{center}
\includegraphics[angle=0,width = 0.45\textwidth]{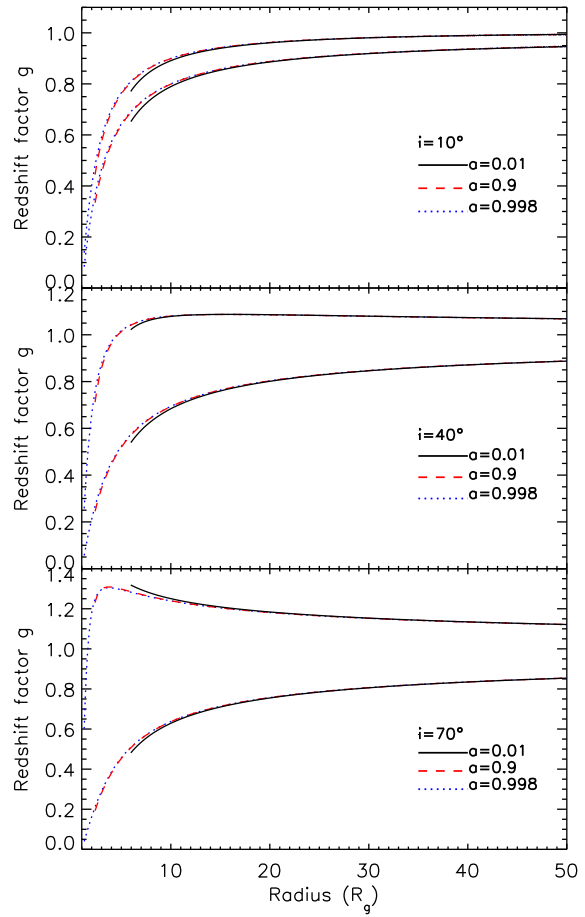}
\caption{The maximal and minimal redshift factor $g$ for a given range of radius at different inclination angles, where the Keplerian rotation velocity is taken. The solid, dashed and dotted lines correspond to black hole spin of 0.01, 0.9 and 0.998, respectively.
The upper, middle, lower panels correspond to the inclination angles of $10^\circ$ , $40^\circ$ and $70^\circ$, respectively.}
\label{fig:gra}
\end{center}
\end{figure}

For the convenience of understanding the results of this paper,
we show the maximal and minimal value of $g$ at different radii along different inclination angles in Figure~\ref{fig:gra}, where the Keplerian rotational velocity is taken for the emitter.
The dotted, dashed and solid lines show the cases of $a=0$, 0.9 and 0.998, respectively.
It can be found that the minimal value of $g$ always increases with radius within 50 $R_g$,
while the maximal value depends on the inclination angle.
For small inclination angles, the maximal $g$ monotonically increases with the radii;
for medium inclination angles about $40^\circ$, it increases with the radius quickly to a value slightly greater than one, and then remains almost unchanged;
while for high inclination angles, it first increases with the radius quickly, then it achieves the maximum at a critical radius, and then decreases with the radius.
For even larger radii, as the gravitational redshift becomes weaker and the azimuthal velocity decreases, both the maximal and minimal $g$ approach unity.

For a narrow ring at a given radius, the width of the line is determined by the maximal and minimal value of $g$, the red and blue peaks are determined by the integration of $g$ over azimuthal angle.
For a ring or subdisc of a certain width, the width of the line is determined by the maximal and minimal value of $g$ over the radial range, and the red and blue peaks are determined by the integration of flux over the azimuthal angle and the radius.
Generally, the width of the line is determined by the inner edge, and the peak frequencies are determined by the outer edge.
If the subdisc is within the critical radius of the high-inclination angle case, the profile could be complex as will be shown later.

\subsection{Thermal Doppler broadening}\label{modb}

Thermal motion of the clump gas could also broaden the emission line, as is called thermal Doppler broadening.
The width of the broadened line is written as \citep{RL79},
\begin{equation}
  \Delta{\nu}=\nu_{\rm 0}\sqrt{\frac{2kT}{{m_{\rm_{e}}c^{2}}}},
  \end{equation}
where $\nu_{\rm 0}$ and $\Delta{\nu}$ are the rest-frame frequency of the line and
the width broadened by the Doppler effect, respectively.
The quantity $k$ is the Boltzmann constant, and $m_{\rm_{e}}c^{2}$ is the rest energy of an electron.
The temperature of the clump gas is represented by $T$, which is assumed to be about that of SAD.
Around the ISCO, the widths of lines from BHXBs and AGNs could be up to $\sim$ 0.4 keV and $\sim$ 4 eV, respectively.
We take this effect into account by broadening the rest energy of a Fe K$\alpha$ line photon during emission, i.e., $h\nu_{\rm 0}$=6.4 keV, with a Gaussian distribution \citep{RL79},
\begin{equation}
    I(\nu)=\frac{1}{\Delta{\nu}\sqrt{\pi}}e^{-\frac{(\nu-\nu_{\rm 0})^{2}}{\Delta{\nu}^{2}}}.
\end{equation}
We separate the line into several grids and treat them as independent lines.
Since redshift factor $g$ is irrelevant to the frequency, the observed frequency and flux can be obtained for each line.
Summing up all the broadened lines, we can get the final line profile that takes into account the thermal Doppler effect.

\begin{figure}
\begin{center}
\includegraphics[angle=0,width = 0.5\textwidth]{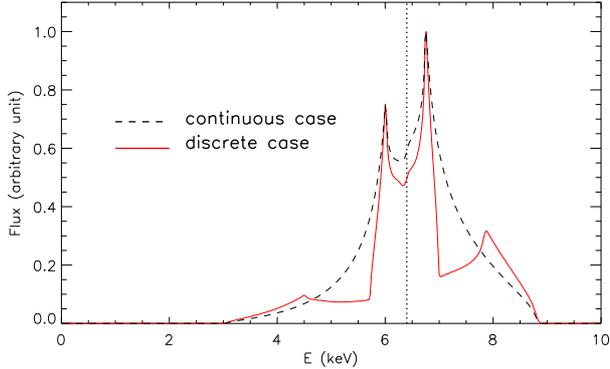}
\caption{Fe K$\alpha$ lines come from continuous and discrete SADs, which are shown with dashed and solid lines, respectively.
For the convenience of comparison, the lines are scaled so that the flux of highest peak is unity.}
\label{fig:cd}
\end{center}
\end{figure}

 \section{Results}\label{res}

There are five parameters for the clumps, i.e. the number of clumps $n$, the covering factor $f$,
the outer boundary of the cold disc $r_{\rm out}$, the truncation radius $r_{\rm t}$,
and the emissivity index $p$.
We investigate the influences of these parameters for different black hole spins $a$ and inclination angles $i$.
As default, the value of clump parameters are taken as
$f=0.1$, $n=1$, $p=3$, $r_{\rm out}=1000$ $R_{\rm g}$, and $r_{\rm t}=50$ $R_{\rm g}$,
where $R_{\rm g}=GM/c^2$.

\subsection{Superimposition of the emission lines}\label{sel}

Generally, instead of double peaks of a SAD, multiple peaks appear due to the superimposition of the lines from different sub-discs,
as shown in Figure~\ref{fig:cd}.
Because the distances of the sub-discs to the central black hole are different,
the peaks of the lines are usually at different frequencies,
and therefore all the peaks can remain after adding up all the lines.

For simplicity, we define $pw$ as the pair of blue/red peaks and wings of one sub-disc.
These $pw$s can be labeled according to the deviation of frequency to 6.4 ${\rm keV}$, or the distance of the sub-disc to the outer cold disc.
For example, the 1st-$pw$ is the pair of peaks and wings that are closest to 6.4 ${\rm keV}$, produced by the outer truncated SAD.

\begin{figure}
\begin{center}
\includegraphics[angle=0,width = 0.5\textwidth]{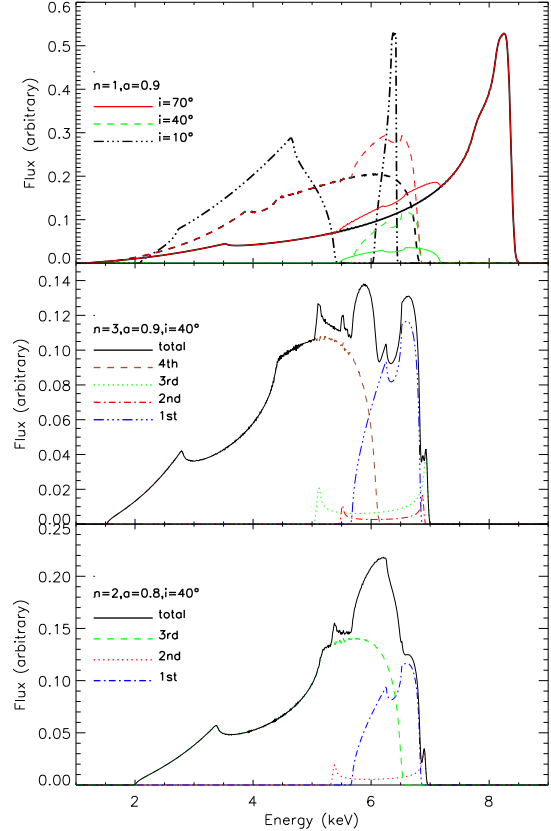}
\caption{Snapshots of some lines. The {\it upper panel} shows the case where the lines of the inner clump and outer SAD separate from each other when the inclination angle is small. The solid, dashed, dash-dotted lines correspond to inclination angles of 70$^\circ$, 40$^\circ$, and 10$^\circ$, respectively. The {\it middle panel} shows the case where a significant fake peak appears. And the {\it lower panel} shows the case where a significant peak forms due to superimposition. }
\label{fig:comp}
\end{center}
\end{figure}

The superimposition of the lines sometimes leads to a misunderstanding, especially when there is a small subdisc around ISCO, as shown in Figure~\ref{fig:comp}.

As shown in the upper panel, given the same narrow clump, line profiles along different inclination angles are very different.
It can be seen that the profile is special when the inclination angle is small.
In the case, the orbital Doppler broadening becomes very weak, as the velocity of the clump is almost perpendicular to the line of sight.
Due to the gravitational redshift effect, the line of the clump separates from the line of the outer cold disc.
As a result, instead of a line with multiple peaks, two separate lines should be observed.

When the blue wing of the inner clump overlaps with the red wing of outer clump/disc, a fake peak can form, as shown in the middle panel.
If the red peak of the outer clump/disc overlaps the blue wing of the inner clump, the red peak will become much more significant, as shown in the lower panel.
In this case, if the resolution is not high enough, the profile looks just like a double-peaked line.

\subsection{The effects of geometric parameters}\label{resg}

\begin{figure}
\begin{center}
\includegraphics[angle=0,width = 0.45\textwidth]{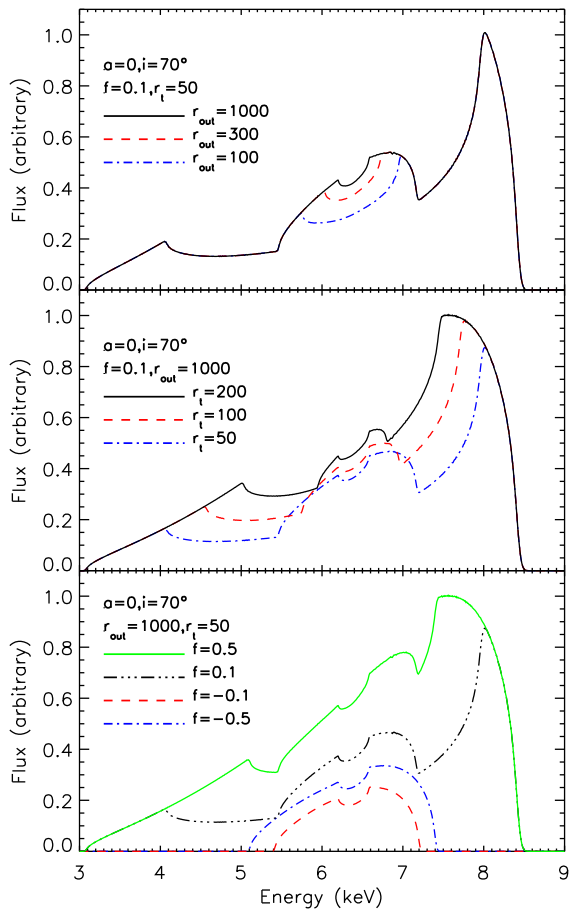}
\caption{This figure shows that the change in the profile of Fe K$\alpha$ line due to the change in $r_{\rm out}$ (upper panel), $r_{\rm t}$ (middle panel) and $f$ (lower panel). The values of spin and inclination angle are represented as 'a' and 'i', respectively. All profiles are scaled to unity for better comparison.}
\label{fig:irrf}
\end{center}
\end{figure}

First we investigate the cases with only a single clump, i.e., $n=1$.
In order to show the influences of the geometric parameters more clearly, Figure~\ref{fig:irrf} is plotted, in which the black hole does not spin and the inclination angle is high.
There are 2 $pw$s in the line profiles, with the 1st-$pw$ being produced by the outer cold disc and the 2nd-$pw$ being produced by the inner clump.
It can be seen that the outer boundary only affects the 1st-$pw$.
When $r_{\rm out}$ increases, and the flux of the 1st-$pw$ also increases.
For given covering fact $f$, both the clump and the outer SAD varies with $r_{\rm t}$, and therefore $r_{\rm t}$ has two effects.
When $r_{\rm t}$ increases, the width of the 1st-$pw$ decreases, while the flux of the 2nd-$pw$ increases.
The covering factor $f$ determines the position and size of the clumps.
For positive $f$, when it increases, the 1st-$pw$ remain the same while the flux of the 2nd-$pw$ increases.
For negative $f$, the clump connect with the outer SAD, and the line is double-peaked.
In this case, the width of the line increases with increasing absolute value of $f$.

\begin{figure*}
\begin{center}
\includegraphics[angle=0,width = 0.9\textwidth]{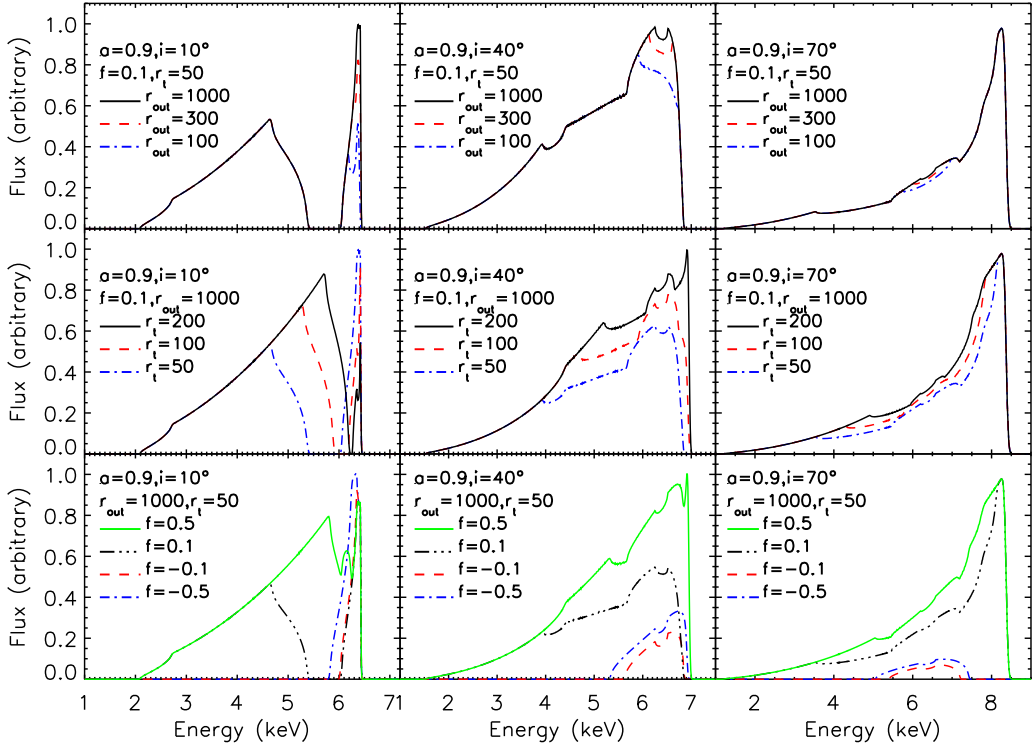}
\caption{The effects of inclination angles $i$ on the line profiles. The value of spin and inclination angle are represented as 'a' and 'i', respectively. All profiles are scaled so that the highest peak is unity.}
\label{fig:thrrf}
\end{center}
\end{figure*}

To better understand the influences of the geometric parameters,
the line profiles at different inclination angles are shown in figure ~\ref{fig:thrrf},
in which the black hole spin is of a typical value, $a=0.9$.
It can be seen that the results are the same qualitatively.

\begin{figure*}
\begin{center}
\includegraphics[angle=0,width = 0.9\textwidth]{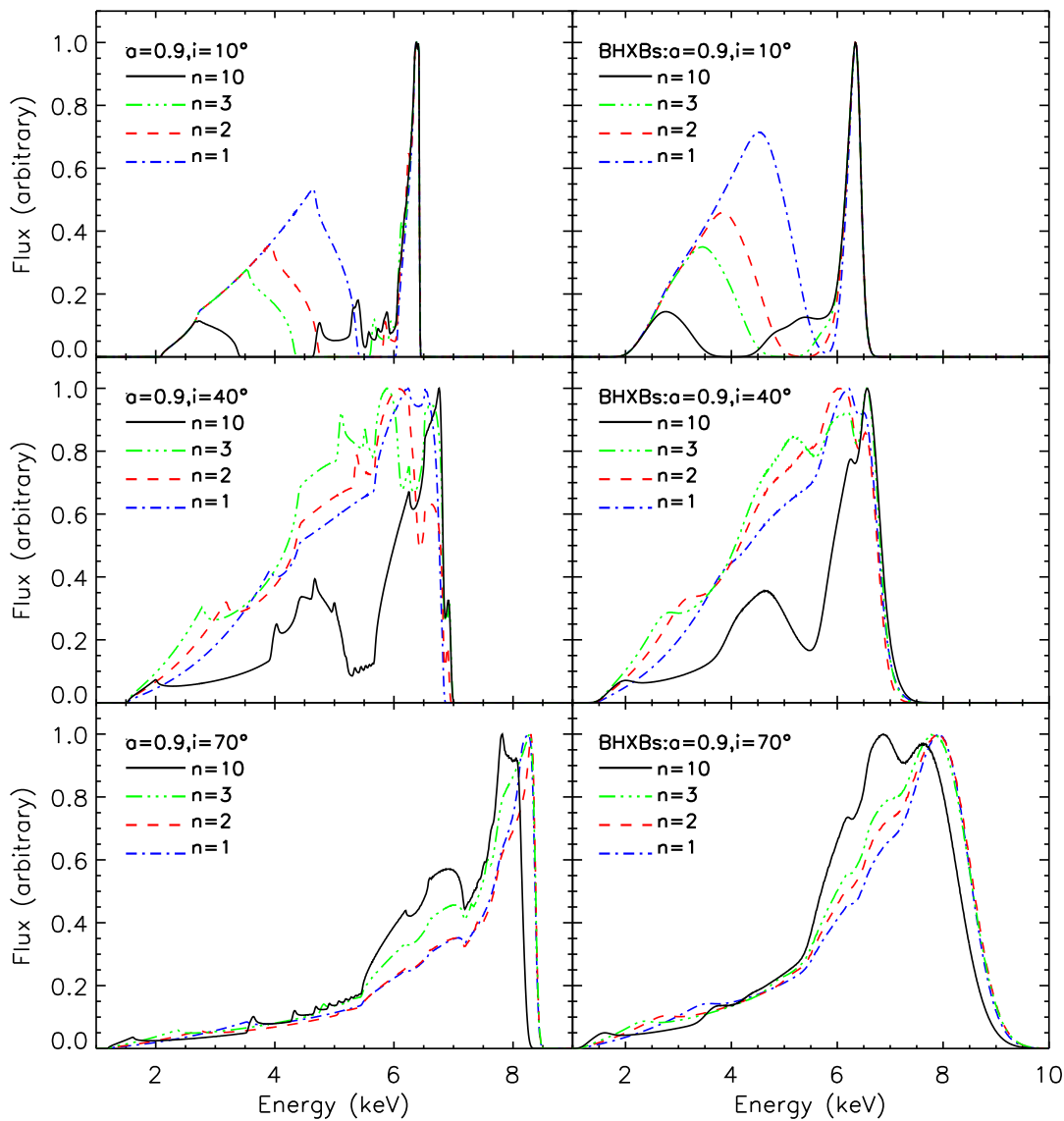}
\caption{The effects of $n$ on the line profiles. {\it Left} and {\it right} panels correspond to the discs of AGNs and BHXBs, respectively. The notations are the same as previous figure.}
\label{fig:ncl}
\end{center}
\end{figure*}

The influences of the number of clumps are shown in Figure~\ref{fig:ncl}.
The left panels show the cases of AGNs with typical mass of $10^{8} M_\odot$,
while the right panels show the cases of BHXBs with typical mass of $10 M_\odot$.
For AGNs, the number of peaks in the line profiles usually increases with $n$, which maybe useful to constrain the number of clumps if high enough resolution is available.
For BHXBs, small peaks are smoothed by thermal Doppler effects.
In this case, it is difficult to constrain the number of clumps,
but it is still possible to identify the existence of clumps as the line profile is not double-peaked.

\subsection{The effects of emissivity index}\label{resp}

\begin{figure*}
\begin{center}
\includegraphics[width = 1.2\textwidth]{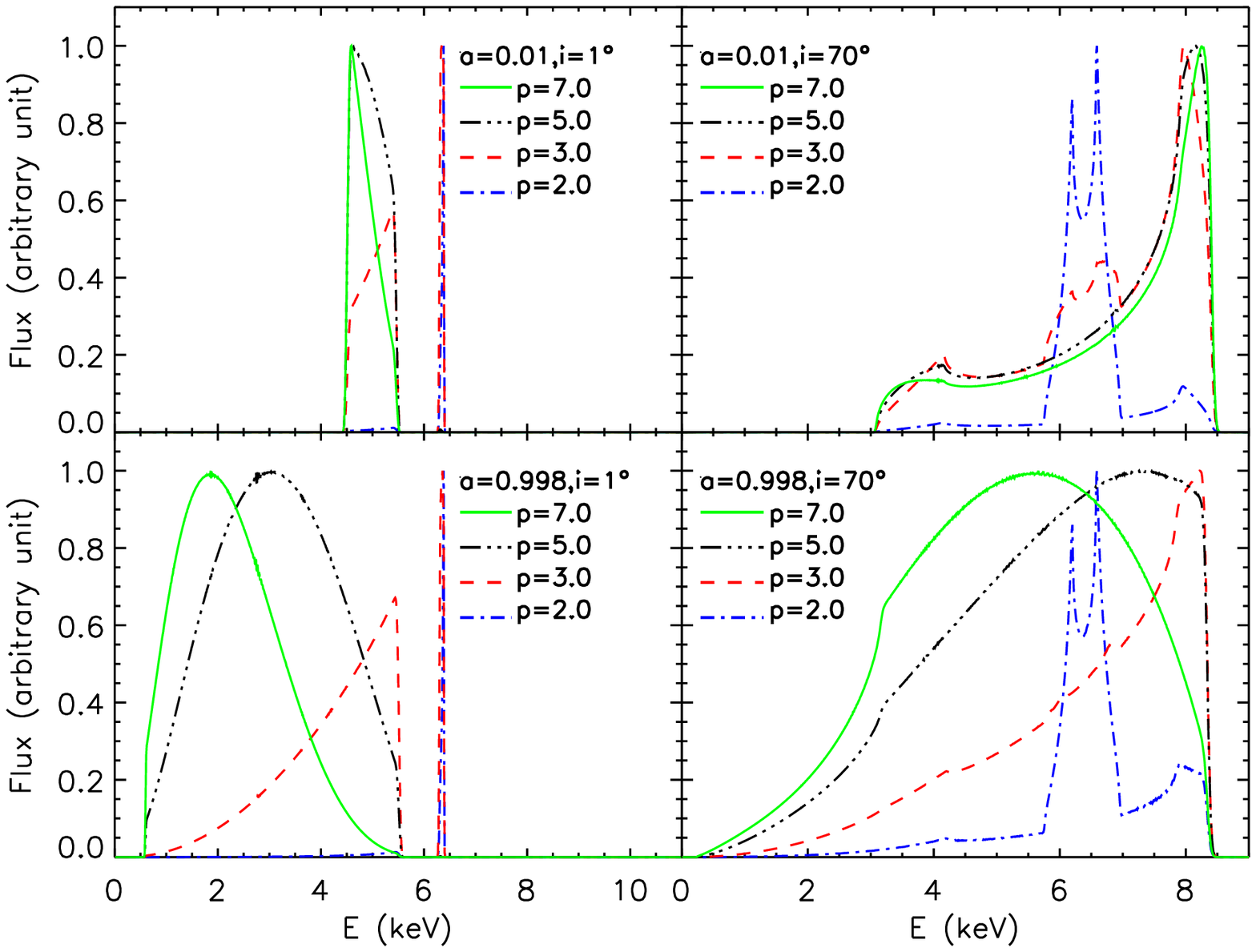}
\caption{The effects of emissivity index on the line profiles. The notations are the same as in previous figure.}
\label{fig:p}
\end{center}
\end{figure*}

The index of emissivity is an important factor in the study of relativistic lines, which describes the distribution of the fluorescent photons or illuminating hard X-rays.
Suppose a point source right above the central black hole, the typical value is 3.
However, some observations show that the value could be much higher.
Here we investigate the influences of emissivity, and the results is shown in Figure~\ref{fig:p}.
In order to show the profiles clearly, each curve is normalized by its highest peak.
Otherwise, the variation of the flux would be so large that it is difficult to observe the peaks when the flux is low.

It can be seen that with the increasing $p$, the contribution of the inner disc becomes more and more important.
When $p\ge 5$, the contribution of the outer cold disc is ignorable.
Consequently, the multi-peak structure can only be found for $p \lesssim 3$.
More calculations show that the critical value of $p$ is not very sensitive to $f$,
which could be understood because of the power-law emissivity.

\section{Discussion}\label{dis}

\subsection{Observations of coexistent broad and narrow Fe lines}\label{disb}

It is interesting that the coexistence of the relativistically broadened line and the narrow line has been found in some Seyfert galaxies \citep[]{B07, N07, C11}.
The narrow lines of some sources are relatively wide, with the width being up to $\sim$ 0.5 keV \citep[]{N07}.
According to our calculations, such width corresponds to a cold reflector lying at the radius of $\sim$ 1000 $R_{\rm g}$ if $i=80^\circ$.
The radius could be even smaller if inclination angle is smaller.
One possible origin of the narrow line is the torus.
However, the typical inner edge of the torus is usually much farther away.
So it is likely that the narrow component is from the truncated SAD instead of the torus.

The broad line component should come from cold clumps or subdiscs in the innermost region that extends to the ISCO.
Since gravitational energy is efficiently released in the inner region,
the radiative efficiency of the cold clumps is high,
and the system is luminous \citep[][]{MM11}.
This could explain the high luminosity of Seyfert galaxies, in addition to the widely believed SAD model.
As for why the hot gas is not cooled down by the soft thermal photons,
it could be due to the the small size of the inner cold disc or energy dissipation in the hot gas.

According to our results, the multi-peak structure is most obvious when the gaps between cold clumps are largest.
So probably there is only one cold clump in these sources, and the clump lies around the ISCO.

Theoretically, inhomogeneous accretion flows could also be found from the sources with only broad iron lines.
As shown in Section~\ref{sel}, when the inclination angle is small, the lines from the clump and the outer cold disc separate from each other.
If we only pay attention to the 1st-$pw$, the signal from the inner clump would be lost.
Moreover, if the red peak of the outer SAD overlaps with the blue wing of the inner clump, the line looks also like a double-peaked broad line.
It would be an interesting test to look for such sources.
Multiple epoch observations should be needed for this purpose.

It should be noted that the solid angle subtended by the truncated SAD from the central X-ray source is much smaller than the solid angle by the torus.
Further studies about the intensities of the lines are still needed.

\subsection{Accretion mode in the intermediate state of BHXBs}

It has been widely accepted that the soft state of BHXBs corresponds to SAD, while the hard state can be well modeled with the truncated SAD by hot accretion flow or corona.
But how the accretion mode transfers between the truncated SAD to the continuous SAD that extends to the ISCO remains unclear.
As for the hard-to-soft transition, there are three possible processes,
the gradual receding of the truncation radius,
formation and expansion of an inner cold disc around ISCO,
formation and expansion of many cold clumps inside the hot gas.
As mentioned previously, in our model, these possibilities correspond to the cases with $n=1$ and $f<0$, $n=1$ and $f>0$, and $n>1$, respectively.

Based on our results, it is theoretically possible to infer the accretion mode during transition from the evolution of the line profiles.
If the truncation radius gradually recedes, the line would be double-peaked all the time, and only the width would change during evolution;
If an interrupted disc or clumps come into form, multi-peak structure would appear.
If a number of clumps are formed, it is possible to distinguish from the interrupted discs from the line profile.
But if the number of clumps is small, it would be difficult to determine whether $n=1$ or not.

\subsection{Toroidal extension of the clumps}
In our model, the clumps are simplified as a series of rings.
In fact, the clumps could take the form of arcs instead of rings.
Especially, for the clumps that are far from the central black hole,
the tidal force they suffered is small,
and their shapes are not elongated.
For these clumps of limited toroidal extension,
we can treat them as short arcs.

If the observation time is shorter than the dynamic timescale of the clumps,
it is difficult to predict the line profile, since the azimuthal positions of the clumps are random.
But if the observation time is much longer than the dynamic timescale of the clumps,
on average, the line profile of a clump is the same as that of a ring at the same radius.
In this case, the line profile of the accretion flow could be estimated from our previous results.

If the toroidal extension of the clump is limited,
the surface available for reflection becomes smaller than the ring of the same radius.
In contrast, the emission from the outer truncated SAD remains the same.
As a result, compared to our previous results, the relative strength of the component from the outer SAD becomes stronger.
The variation of the relative strength could be calculated from the toroidal extension of the clumps.
At a given radius, if there are more than one clumps, the variation is proportional to the total central angle that all the arcs of the clumps subtend to the black hole.

Theoretically, the time-averaged line profile can be calculated for any distribution of the toroidal extension over the radius.
But physically, the distribution of the clump size is still unclear.
One possibility is that the toroidal extension is a constant with fluctuation.
Whatever, the clump size leaves an uncertainty in modeling the line profile.

As shown in the previous section, only when the emissivity index is $p \lesssim 3$, does the multi-peak structure remain.
When we take into account the toroidal extensions of the clumps,
the critical value of $p\sim 3$ can be higher.
If we assume the central angles from the clumps to the center are constant,
when it changes from $2\pi$ to $0.01\pi$, the critical value changes from $\sim 3$ to $\sim 5$ for fast spin black holes,
and from $\sim 5$ to $\sim 7$ for slowly rotating black holes.

If the radial velocity is comparable to the azimuthal velocity,
the clumps will take the form of spirals instead of rings or arcs.
Some works explored the instantaneous relativistic Fe K$\alpha$ line emitted from the spirals \citep[]{HB02,FT04}.
They found that, except for the multi-peak feature shown in this paper, there are many spiky peaks overlaying the multiple peaks.

\subsection{Caveats}\label{disd}

An important simplification of the present paper is that we assume the iron atoms to be neutral, ignoring the ionization state of the cold gas.
In fact, due to the intensive hard X-ray illumination, the disc surfaces should be highly ionized.
This will significantly affect the emergent spectra in the energy band we are interested in, i.e., $\sim 2-8$ keV.
Because the disc surfaces are dominated by partially ionized iron,
the intrinsic profile should include a series of lines produced by different iron ions \citep[]{R99,GK10}.
Secondly, the fully ionized light elements at the surface could form a layer of large Thomson optical depth, which would scatter the neutral iron line photons and would produce a Compton shoulder \citep[]{M91,W03,MT07}.
These effects could be included by considering the balance between ionization and recombination, and could be studied in future works.

Before being observed, the line photons have to cross the surrounding hot plasma.
The consequent absorption and inverse Compton scattering can make the problem more complex.
However, the influences of the ambient hot gas are not included in this paper.
So the present results are correct only for the cases when the optical depth of the hot gas is small.
Considering related research, the optical depth should be less than 0.1 for gas temperature of $\sim 10^9$K \citep[e.g.][]{Wilkins15}.

Our results are about the time-averaged spectra, the instantaneous line is out of the scope of this paper.
However, this variability of Fe lines is becoming an important method to study the accretion flow near the BH \citep{M07}.
For the model of an orbiting flare over SAD, \citet{I04} show a saw-tooth pattern.
Some observations have been done on this variation pattern \citep[][]{Turner06,DeM09,Nardini16}.
It will be interesting to investigate the variation pattern of clumps in future works.

\section{Conclusions}\label{con}
In this work, the relativistic Fe K$\alpha$ line from an inhomogeneous disc is calculated.
If the emissivity is not very high, multiple peaks appear.
The size and position of the clumps affect the profiles.
Cautions should be taken when fitting the lines,
especially when a small clump exists in the innermost region.

\section*{Acknowledgements}

The authors thank the anonymous referee for the valuable comments.
We are grateful to Chris Talmage for her help with the English.
This work was supported by the National Natural Science Foundation of
China under grants U1531130, 11333004, 11525312, 11443009, and 11703064, and the Fundamental Research Funds for the Central University under grants 20720150024.
YPL was also sponsored in part by Shanghai Sailing Program (No. 17YF1422600).

% Don't change these lines
\bsp	% typesetting comment
\label{lastpage}
\end{document}